\documentclass[12]{article}
\usepackage{epsfig}
\usepackage{amsmath,amssymb}
\usepackage{textcomp}

\newcommand{\be}{\begin{equation}}
\newcommand{\ee}{\end{equation}}

\newcommand{\LL}{\mathcal{L}}

\title{Electroweak Theory without a Higgs potential: Radiative Effects}
\author{Srijit Bhattacharjee and Parthasarathi Majumdar \\{\it Theory Division,
  Saha Institute of Nuclear Physics} \\ {\it AF/1 Bidhannagar, Kolkata 700022, INDIA.}}

\begin{document}

\maketitle

\begin{abstract}

We examine the one loop effective potential in a recently proposed (by Chernodub
et. al.) alternative approach to mass
generation in the Higgs-gauge sector of the electroweak theory, which does not make
use of a classical Higgs potential. We show that the interpretation given by
these authors, of the
Higgs boson as the conformal degree of freedom in a background conformal gravity
theory, is invalidated because genuine one loop radiative effects cancel the
local functional measure of the Higgs field taken to be the basis of this
interpretation. Functional evaluation of the one loop effective Higgs
potential leads to a minimum away from the origin, thereby providing yet
another radiative mass generation scheme for weak gauge bosons. In arriving at
the one loop effective potential we make use of the gauge free formulation of
electrodynamics introduced by us earlier, which obviates any need for gauge
fixing.   

\end{abstract}

\section{Introduction}

Recently a new approach to the Higgs phenomenon of mass generation in
the gauge boson-Higgs boson sector of standard electroweak gauge
theory has been proposed \cite{Chernodub:2008rz}, \cite{Faddeev:2008qc},
\cite{Faddeev:2006sw}, \cite{Chernodub:2007bz} together with a novel
interpretation of the Higgs scalar degree of freedom. Beginning with a
standard $SU(2) \times U(1)$ gauge theory augmented by a Higgs action, but {\it
  without} a Higgs potential, field redefinitions are performed whereby all
fields are rendered completely {\it inert} under $SU(2)$ gauge transformations.\footnote{This
is perhaps related to the gauge-free electrodynamics \cite{Majumdar:2009yw} proposed by us
elsewhere, generalized to the context of a non-Abelian gauge theory with Higgs
couplings.} As a result of field redefinitions, the theory has a
remaining Abelian gauge invariance under
$U(1)_{em}$ with the photon being left over as the only massless gauge
field. The gauge neutral modulus of the Higgs field couples to the $W$ and $Z$
bosons in the standard manner and can indeed provide masses to them if it
picks up a vacuum expectation value. Unlike in the standard Higgs mechanism
which employs a (unstable) Higgs potential for this purpose, the Higgs degree
of freedom here is given a novel interpretation: it is the conformal degree of
freedom in a conformally flat background gravity theory, such that its vacuum
value is fixed by its large distance (cosmological) behaviour. This interpretation
derives from the altered functional measure for the Higgs modulus field
because of field redefinitions: it is {\it local} in nature. Indeed, this is a
fascinating alternative to the standard electroweak Higgs mechanism if it
survives quantization. The possibility that this may unfortunately {\it not}
be the case has already been considered in \cite{Ryskin:2009kw}. In this paper, we
explicitly consider perturbative radiative
effects in the entire scenario to study this question in more detail. We consider
the one loop effective potential of the theory and find that the local terms
in the functional measure in fact do {\it not} survive quantization : they are
exactly cancelled by genuine radiative terms arising from the functional
evaluation of the one loop effective potential. The novel interpretation is
thus subject to modification in a full quantum field theory treatment.

Rather than attempting to resurrect the interpretation, we probe the question
as to whether an effective Higgs potential might emerge {\it radiatively}.  The one loop
effective potential we compute indeed has a
minimum away from the origin, determined by the renormalization scale and
the gauge couplings. There is thus adequate structure at the
quantum level of the theory to generate gauge boson (and Higgs) masses in terms
of its parameters. We compare the results of the Higgs-vector mass
ratio obtained here to that obtained in the standard Coleman-Weinberg approach
\cite{Coleman:1973jx} where conventional Higgs mechanism is implemented radiatively.

The plan of the paper is as follows: in section 2 the field redefinitions
employed in \cite{Faddeev:2008qc} that render all fields free of $SU(2)$ gauge
transformations are discussed briefly. These are important because they
enable us to work with $SU(2)$ invariant fields, thereby obviating the need to
perform $SU(2)$ gauge fixing. We extend these field redefinitions to the
$U(1)_{em}$ invariant residual theory as well, following our earlier work
\cite{Majumdar:2009yw}, so that all fields are rendered
manifestly $U(1)$ invariant, thereby avoiding any gauge fixing at all.
This is followed in section 3 with a computation of the
one loop effective Higgs potential, using the same local measure discerned in
\cite{Faddeev:2008qc}. It is shown how the effects of the local measure are cancelled by
radiative terms in the effective potential. Appropriate renormalizations are
performed to yield a renormalized one loop effective Higgs potential. In
section 4, it is shown that the effective Higgs potential possesses an
absolute minimum away from the origin in Higgs field space, and discuss the
spectrum of particles around this minumum. The concluding section (5) contains
a discussion of our results.     

\section{New variable form of $SU(2) \times U(1)$ theory}

This section is entirely based on \cite{Faddeev:2008qc}. The gauge-Higgs sector of the
standard electroweak theory contains the $SU(2)$ 
gauge field ${\bf B}_{\mu} = B_{\mu}^a t^a, a=1,2,3$, the $U(1)$ gauge field $Y_{\mu}$ and the Higgs $SU(2)$
doublet $\Phi$, transforming under $SU(2) \times U(1)$ gauge tranformations as
\begin{eqnarray}
{\bf B}_{\mu} \rightarrow {\bf B}^{(\Omega)}_{\mu} &=& \Omega {\bf B}_{\mu}
\Omega^{-1} - \partial_{\mu} \Omega \Omega^{-1} \nonumber \\
Y_{\mu} \rightarrow Y_{\mu}^{(\omega)} &=& Y_{\mu} + \partial \omega \nonumber  \\
\Phi \rightarrow \Phi^{(\Omega)} = \Phi \Omega &,& \Phi \rightarrow
\Phi^{(\omega)} = \Phi \exp i \omega . \label{gtr}
\end{eqnarray} 
 
The Lagrange density for this sector of the electroweak theory is
\begin{equation}
    \LL= \left(\nabla_{\mu}\Phi \nabla^{\mu}\Phi\right) - \frac{1}{4g^{2}}tr
	{\bf B}^2_{\mu\nu} -\frac{1}{4g'^{2}} Y_{\mu\nu}^{2} , \label{lagew}
\end{equation}

    where
\begin{align*}
    \nabla_{\mu}\Phi =& \partial_{\mu} \Phi +\frac{i}{2} Y_{\mu} \Phi
	+ B^{a}_{\mu} t^{a} \Phi \\
    B^{a}_{\mu\nu} =& \partial_{\mu} B^{a}_{\nu} - \partial_{\nu} B^a_{\mu}
	+ \epsilon_{abc} B^{b}_{\mu} B^{c}_{\nu} \\
    Y_{\mu\nu} =& \partial_{\mu} Y_{\nu} - \partial_{\nu} Y_{\mu}
\end{align*}
with
$ t^{a} = \frac{i}{2} \tau^{a}$,
$ \tau^{a} $ Pauli matrices,
$ g $ and $ g' $ are the coupling constants.

The essential feature of this approach is the `polar decomposition' of the complex scalar doublet into two parts.
\begin{equation}
    \Phi ={1 \over {\sqrt{2}}} \rho \chi ,
\end{equation}
where, $\rho$ is a real positive scalar (modulus) field is completely gauge-inert, while the `phase'
part $\chi$ carries all the gauge transformation properties of $\Phi$. Now,
one introduces the matrix
\begin{equation}
    g=
\begin{pmatrix}
    \chi_{1} & - \bar{\chi}_{2} \\
    \chi_{2} & \bar{\chi}_{1} ~\label{gee}
\end{pmatrix}
\end{equation}
with a normalisation $(\chi,\chi) = \bar{\chi}_{1} \chi_{1} + \bar{\chi}_{2}
\chi_{2} = 1$ (which defines the group manifold of $SU(2)$), it is easy to
verify \cite{Faddeev:2008qc} that $g$  is unimodular and unitary. It stands to reason
that $g \in SU(2)$ so that under an $SU(2)$ gauge transformation 
\begin{equation}
g \rightarrow g^{(\Omega)} = \Omega g . \label{geeg}
\end{equation}
However, since $\chi_i$ and ${\bar \chi}_i$ have different weak hypercharges,
under a $U(1)$ gauge transformation, $g \rightarrow g^{(\omega)}=g e^{i\omega \tau_{3}}$

The covariant derivative of $g$ is given by 
\begin{equation}
    \nabla_{\mu} g = \partial_{\mu} g + \frac{i}{2} Y_{\mu} g \tau_{3}
	+ {\bf B}_{\mu} g ~. \label{covg}
\end{equation}

Defining the new Yang Mills triplet ${\bf W}_{\mu} = W_{\mu}^a t^a$ as
\begin{equation}
{\bf W}_{\mu} \equiv g^{\dag} \left( {\bf B}_{\mu} + \partial_{\mu} \right) g
\end{equation}
it is easy to see that under an $SU(2) \times U(1)$ gauge transformations, 
\begin{eqnarray}
{\bf W}_{\mu}^{(\Omega)} &=& {\bf W}_{\mu} \nonumber \\
{\bf W}_{\mu}^{(\omega)} &=& e^{-i\omega \tau_3} {\bf W}_{\mu} e^{i \omega
  \tau_3} + i\tau_3 \partial_{\mu}\omega . ~\label{ginv} 
\end{eqnarray}
These fields ${\bf W}_{\mu}$ are thus explicitly $SU(2)$ gauge invariant, even
though they have nontrivial $U(1)$ gauge transformations. One defines the
linear combinations
\begin{eqnarray}
Z_{\mu} &\equiv& Y_{\mu} + W_{\mu}^3 \nonumber \\
A_{\mu} & \equiv & {1 \over g^2 + g'^2}~(g'^2 W_{\mu}^3 - g^2 Y_{\mu})
~, \label{neut}
\end{eqnarray}
where, the vector field $Z_{\mu}$ is manifestly $SU(2) \times U(1)$ invariant,
and the $A_{\mu}$ field transforms under $U(1)$ as $A^{(\omega)} = A_{\mu}
-2 \partial_{\mu} \omega$. The charged combinations $W_{\mu}^{\pm} \equiv
W^1_{\mu} \tau_1 \pm W^2_{\mu} \tau_2$ are $SU(2)$ gauge invariant, but carry
indicated charges under $U(1)_{em}$ gauge transformations under which
$A_{\mu}$ transforms as the photon field, with the electronic charge being
$e^{-2} \equiv g^{-2} + g'^{-2}$.     

The entire gamut of field redefinitions leave only a $U(1)_{em}$ gauge theory
with the photon field being the sole gauge connection; the Yang Mills
connections have been rendered entirely gauge free under $SU(2)$ gauge
transformations, and behave as charged (or neutral) vectorial matter fields under
electromagnetism. The theory is described by the Lagrange density
\begin{align}
    \LL =&{1 \over 2} \partial_{\mu} \rho \partial^{\mu} \rho +\frac{\rho^{2}}{8}
	(Z_{\mu}^{2}+W_{\mu}^{+}W^{\mu,-}) 
    - \frac{1}{4g^{2}} (\nabla_{\mu}W_{\nu}^{+}-\nabla_{\nu}W_{\mu}^{+})
	(\nabla^{\mu}W^{\nu,-}-\nabla^{\nu}W^{\mu,-}) \nonumber \\
    -& \frac{1}{4(g^{2}+g'^{2})} Z_{\mu\nu}^{2} -
	\frac{1}{4e^{2}} A_{\mu\nu}^{2} 
    - \frac{2}{4g^{2}}~H_{\mu\nu}(A^{\mu \nu}+e^2 Z^{\mu \nu})
	-\frac{1}{4g^{2}}H_{\mu\nu}^{2} , \label{lagr}
\end{align}
 where 
\begin{eqnarray}
    Z_{\mu\nu} &=& \partial_{\mu}Z_{\nu} -\partial_{\nu}Z_{\mu} \nonumber \\
    A_{\mu\nu} &=& \partial_{\mu}A_{\nu} -\partial_{\nu}A_{\mu} \nonumber \\
    W_{\mu\nu}^{3} &=& \partial_{\mu} W_{\nu}^{3} - \partial_{\nu} W_{\mu}^{3}
    \nonumber \\
    H_{\mu\nu} &=& \frac{1}{2i} (W_{\mu}^{+}W_{\nu}^{-}
    -W_{\mu}^{-}W_{\nu}^{-}) \nonumber \\
    \nabla_{\mu} W_{\nu}^{\pm} &=& \partial_{\mu}W_{\nu}^{\pm} \pm 
	i W_{\mu}^{3} W_{\nu}^{\pm} \label{curv} .
\end{eqnarray}

The question is : does this theory generate a mass for the $W_{\mu}^{\pm}$, the $Z_{\mu}$
and the $\rho$ fields, as is achieved in the standard Higgs mechanism by means
of a Higgs potential with degenerate minima ? In other words, what is the
scale of the vacuum expectation value $\rho$ here, since there is no Higgs
potential to generate that scale ? In \cite{Chernodub:2008rz} the Higgs modulus
field $\rho$ is interpreted, because of the appearance of the local $\rho^2
{\cal D} \rho$ factor that appears in the partition functional integral, as
the conformal factor of a background conformally flat spacetime. The vacuum
value of $\rho$ is related to its asymptotic value in this spacetime, and is
supposed to be determined cosmologically. Excitations around this vacuum value
are of course to be interpreted as a new massless scalar field. So, a new
perturbative mechanism to produce vector boson masses becomes available, as an alternative to
the standard Higgs mechanism. The issue is: does this mechanism survive
quantization ?

Even though the theory has a residual $U(1)_{em}$ gauge invariance, one can
rewrite it explicitly in terms of entirely {\it gauge free} variables \cite{Majumdar:2009yw}, so that
no gauge fixing is at all necessary to evaluate the partition function. We
begin by a radial decomposition of the charged weak vector boson fields
\begin{eqnarray}
W_{\mu}^{\pm} = w_{\mu} \exp \pm i\theta^{(\mu)}~, no~sum~on~\mu \label{wrad}
\end{eqnarray}   
which implies that under $U(1)$ gauge transformations
\begin{eqnarray}
[w_{\mu}]^{(\omega)} = w_{\mu}~,~[\theta^{(\mu)}]^{(\omega)} = \theta^{(\mu)}
+2\omega ~. \label{wgau}
\end{eqnarray} 
One can think of $w^{\mu}$ as the component of the charged vector boson
carrying only the {\it spin} while $\theta^{(\mu)}$ is the {\it charge} mode,
thus affecting a `separation of the charge and spin modes'. It follows that
\begin{eqnarray}
\nabla_{\mu} W_{\nu}^{\pm} = \left[\partial_{\mu} w_{\nu} \pm
iw_{\nu}\left( \partial_{\mu}\theta^{(\nu)} - A_{\mu} -{e^2 \over
g'^2}Z_{\mu}\right) \right] e^{i\theta^{(\nu)}}
\end{eqnarray} 
The only quantity sensitive to $U(1)_{em}$ gauge transformations is the phase factor; the
gauge transformation parameter cancels between the $\theta$ and $A$ fields
within the parantheses. However, all fields can now be expressed explicitly in terms of
entirely gauge free fields except the phase factor which indeed must carry the
full burden of gauge transformations, through the field redefinitions
\cite{Majumdar:2009yw}. 

\begin{eqnarray}
\Theta^{(\mu)} &\equiv& \theta^{(\mu)} - 2 {\rm a} ~\nonumber \\
{\bf A}_{\mu} & \equiv & A_{\mu} - 2 \partial_{\mu} {\rm a} ~, \label{redef}
\end{eqnarray}
where, ${\rm a}(x) \equiv \int d^4x' G(x-x') \partial' \cdot A(x')$ is a scalar
field giving the longitudinal mode of $A_{\mu}(x)$ with $G(x-x')$ being the
Green's function for the d'Alembertian. As a result of these field
redefinitions, the kinetic energy of the charged
vector bosons (and indeed the entire Lagrange density) is rendered free of
{\it all} local gauge transformations. The former assumes the form
\begin{eqnarray}
\nabla_{[\mu} W_{\nu]}^+ \nabla^{[\mu} W^{\nu] -} &=& \frac12 \sum_{\mu, \nu
  =0}^3 \{
[ \partial_{\mu} w_{\nu} \partial^{\mu} w^{\nu} +  w_{\nu} ({\tilde A}^{(\nu)}_{\mu} + {e^2 \over
  g'^2} Z_{\mu}) w^{\nu}( {\tilde A}^{\mu (\nu)}  + {e^2 \over g'^2} Z^{\mu})  ] \nonumber \\
&-& \cos \Theta^{(\mu \nu)} [ \partial_{\mu} w_{\nu} \partial^{\nu} w^{\mu} +
w_{\nu} ( {\tilde A}^{(\nu)}_{\mu} +{e^2 \over g'^2}Z_{\mu}) w^{\mu} ({\tilde
  A}^{(\mu) \nu} +{e^2 \over g'^2} Z^{\nu} ) ] \nonumber \\
&-& \sin \Theta^{(\nu \mu)} [ \partial_{\nu}w_{\mu} w^{\nu} ({\tilde A}^{(\nu)
  \mu} + {e^2 \over g'^2}Z^{\mu} ) ] \} ~\label{nabs}
\end{eqnarray}
where, ${\tilde A}^{(\mu)}_{\nu} \equiv {\bf A}_{\nu} - \partial_{\nu}
\Theta^{(\mu)}$ and $\Theta^{(\mu \nu)} \equiv \Theta^{(\mu)} - \Theta^{(\nu)}$ 

Since the $W^{\pm}$ carry electric charge $\pm1$, one can in fact choose the phases
$\Theta^{(\mu)}$ to be the same, independent of the $\mu$, without any loss of
generality. With this choice, eq. (\ref{nabs}) simplifies considerably
\begin{eqnarray}
\nabla_{[\mu} W_{\nu]}^+ \nabla^{[\mu} W^{\nu] -} &=& w_{\mu \nu}^2 + \frac 12
w^2 \left( {\bf A} - \partial \Theta + {e^2 \over g'^2} Z \right)^2
\nonumber\\
&-& \frac12 \left [w \cdot \left( {\bf A} - \partial \Theta + {e^2 \over g'^2}
Z \right) \right]^2 ~, \label{nabs0}
\end{eqnarray}
where, $w_{\mu \nu} \equiv 2\partial_{[\mu} w_{\nu]}$. This equation exhibits
the $U(1)_{em}$ gauge freedom of the fields manifestly, and also shows
explicitly the coupling of the charged vector boson modes to the physical $U(1)$ photon vector potential.

\section{One Loop Effective Potential}

We now turn to the question of the perturbative quantum behaviour of the theory.
To study this question, we consider the one loop effective potential of the
theory given by (\ref{lagr}), and investigate if it has a nontrivial minimum
driven by infrared instabilities as in the Coleman-Weinberg mechanism
\cite{Coleman:1973jx}. We do not use the new interpretation given in \cite{Chernodub:2008rz}. The issue
then amounts to investigating the possibility of radiative generation of a
Higgs potential (not just a mass term), with a self-coupling {\it determined}
in terms of the gauge couplings. Since we are interested in only one-loop
calculations we drop all the cubic and quartic interaction terms from the
Lagrangian in (\ref{lagr}) since they do not contribute at the one loop level.
This is easy to see by simply drawing all possible one loop graphs with $\rho$
external lines: none of them have the vertices that are being discarded here.
One is thus dealing with the truncated Lagrangian relevant for one-loop calculations,
\begin{align}
    \LL_{trun} =&{1 \over 2} \partial_{\mu} \rho \partial^{\mu} \rho +\frac{\rho^{2}}{8}
	(Z_{\mu}^{2}+w_{\mu}^2) 
    - \frac{1}{4g^{2}} w_{\mu \nu}^2 \nonumber \\
    -&\frac{1}{4e^{2}} {\bf A}_{\mu\nu}^{2} -\frac{1}{4(g^{2}+g'^{2})}
    Z_{\mu\nu}^{2} ~, \label{lsimpl}
\end{align}
where, the physical photon field is divergenceless $\partial \cdot {\bf A} =0$
as discussed in \cite{Majumdar:2009yw}. 

The generating functional of all Feynman graphs is given by 
\begin{align*}
Z[J, {\bf J}_A, {\bf J}_Z, {\bf J}^+, {\bf J}^-]=&\int d\mu ~ \delta[\partial
\cdot {\bf A}]~ \exp {i\int d^4x\,
  \LL_{trun}}\\ .& \exp { i\int d^4x (J\rho +{\bf J}_{\bf A}.{\bf A} +{\bf J}_Z.Z +{\bf
J}_w.w + J_{\Theta} \Theta)}
\end{align*}
where the measure \[d\mu=Det\rho^2 {\cal D}\rho^2{\cal D}{\bf A} {\cal D} a {\cal D}Z{\cal
D}w{\cal D} \Theta {\cal D}g\] where ${\cal D}g$ is the $SU(2)$ group volume
and ${\cal D}a $ is that of the $U(1)_{em}$. Since the action is manifestly
independent of the fields characrterizing these group volumes, they can be
factored out of the functional integral and discarded as irrelevant
multiplicative factors. Note however the {\it local} measure associated with
the Higgs field $\rho$; this implies that $\rho$ is not an usual scalar field,
as pointed out in \cite{Chernodub:2008rz}, \cite{ Faddeev:2008qc}. It appears to
behave like a {\it dilaton} field which might acquire a vacuum value from
cosmological sources. However, this local measure undergoes a precise
cancellation, as we shall show shortly. 

We calculate the one loop effective
potential functionally from the generating functional using saddle-point
method \cite{jackiw74}. Shifting
the positive real scalar field \[\rho \rightarrow \rho_0+\rho\] where
$\rho_0$ is a spacetime constant chosen to be the saddle point, the one-loop effective potential becomes

\begin{align}
   V_{eff}(\rho_0) =& 3i\int \frac{d^4k}{(2\pi)^4}\ln
   (\rho_0)-\frac{i\hbar}{2}\int \frac{d^4k}{(2\pi)^4}\ln
   (-k^2) \nonumber \\-&\frac{i}{2g^{2}}\int
   \frac{d^4k}{(2\pi)^4} \ln Det\left[\eta^{\mu\nu}\left(-k^2+g^{2}\frac{\rho^{2}}{4}\right)+k^{\mu}k^{\nu}\right]
   \nonumber \\-&\frac{i}{2(g^{2}+g'^{2})}\int
   \frac{d^4k}{(2\pi)^4}\ln Det\left[\eta^{\mu\nu}\left(-k^2+(g^{2}+g'^{2})\frac{\rho^{2}}{4}\right)+k^{\mu}k^{\nu}\right]
   \nonumber \\ -&
	\frac{i}{2e^{2}}\int \frac{d^4k}{(2\pi)^4}\ln
        Det\left[-\eta^{\mu\nu}k^2 \right] ~, \label{veff}
\end{align}
where `Det' means a determinant in the functional space. Upon evaluating the eigenvaues of the respective integrals we obtain
\begin{align}
   V_{eff}(\rho_0) =& 3i\int \frac{d^4k}{(2\pi)^4}\ln
   (\rho_0)-\frac{i\hbar}{2}\int \frac{d^4k}{(2\pi)^4} \ln(-k^2) \nonumber\\
-&3i\int \frac{d^4k}{(2\pi)^4}\ln
\left(k^2-g^{2}\frac{\rho^{2}}{4}\right)-i\int \frac{d^4k}{(2\pi)^4}\ln
(\rho_0) \nonumber \\
-&\frac{3i}{2}\int
\frac{d^4k}{(2\pi)^4}\ln\left(k^2-(g^{2}+g'^{2})\frac{\rho^{2}}{4}\right)
-2i\int \frac{d^4k}{(2\pi)^4}\ln \rho_0~. \label{vef}
\end{align}

One significant outcome of the above expression is that the Jacobian contribution from the functional
measure discerned in \cite{Chernodub:2008rz} (the 1st term in (\ref{vef})) is exactly cancelled by
two terms coming from the neutral and charged vector boson operators (the fourth
and seventh terms respectively in (\ref{vef})). Another fact of this result is
photon part of the Lagrangian doesn't contribute to the one-loop effective
potential since it is not coupled to any other field of the theory. Thus on
integration it will give an irrelevant infinite constant which we have
subtracted off. After introducing renormalizing
counterterms and Wick rotating the contour of integration, the effective potential becomes

\begin{align*}
V_{eff}(\rho_0)=\,\,& \frac{B}{2}\rho_0^2 \,+\,\frac{C}{4!}\rho_0^4 \nonumber \\
\,\,+&\frac{3}{2}\int \frac{d^4k_E}{(2\pi)^4}
\,\ln\left(k_E^2+(g^{2}+g'^{2})\frac{\rho^{2}}{4}\right) \,+\, 3\int
\frac{d^4k_E}{(2\pi)^4} \ln \left(k_E^2+g^{2}\frac{\rho^{2}}{4}\right)~, 
\end{align*} 
where B, C are the usual mass and coupling-constant renormalization counter
terms. The counterterms are determined using the renormalization
scheme \be \frac{d^2 V}{d\rho_0^2}\Bigg|_{\rho_0=M}=0 \label{renoMcw} \ee and
\be\frac{d^4 V}{d\rho_0^4}\Bigg|_{\rho_0=M}=0 \label{renocCW}\ee
where the parameter $M$ serves as a scale of the theory. The renormalized
effective one loop Higgs potential is now given by 

\begin{align}
V_{eff}(\rho_0)= \,&\nonumber\frac{27(g^{2}+g'^{2})^2M^2\rho_0^2}{512\pi^2}+\frac{27g^4M^2\rho_0^2}{256\pi^2}\\
+ \,&\left(\frac{3(g^{2}+g'^{2})^2\rho_0^4}{1024\pi^2}+\frac{3g^4\rho_0^4}{512\pi^2}\right)\left(\ln\frac{\rho_0^2}{M^2}
-\frac{25}{6}\right)~,     \label{effPwh}
\end{align}

\begin{figure}
    \begin{center}
      \resizebox{120mm}{!}{\includegraphics{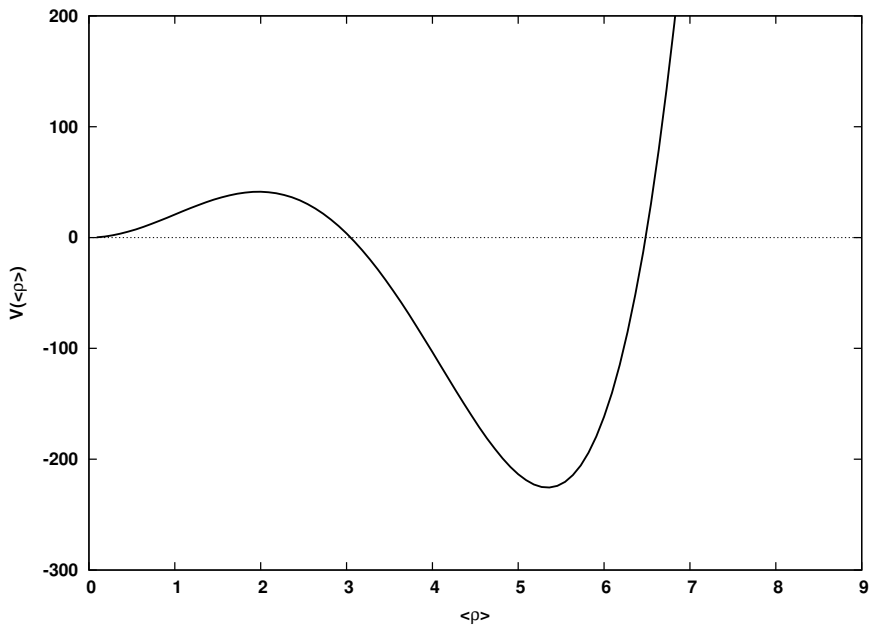}}
      \caption{Plot of the Effective Potential as a function of $\frac{\langle{\rho}\rangle}{M}$.}
      \label{test}
    \end{center}
  \end{figure}

Henceforth we drop the subscript $0$ on $\rho_0$. The plot of effective
potential (Fig. [\ref{test}]) shows that it has three extrema in the
physically interesting region i.e. in the region where $\langle{\rho}\rangle >0$.
Apart from a local minima at the origin the potential posseses a
maxima at about $\langle{\rho}\rangle\simeq 1.98M$. The true minimum is around
$\langle{\rho}\rangle \simeq 5.34M$ for which $\log (\langle{\rho} \rangle /
M) \simeq 1.67$. 

The mass generated for the $W^{\pm}$ bosons is
$(1/2)g\langle \rho \rangle$, while for the $Z$ bosons is
$(1/2)(g^2+g'^2)^{1/2} \langle \rho \rangle$. Thus, since $\langle \rho
\rangle = 246 Gev$ reproduces the observed $W$ and $Z$ boson mass spectrum,
this implies that one must make the choice $M \sim 46$ Gev 
This also corresponds to a $\rho$ boson mass, computed as the
curvature of the effective potential at the absolute minimum, 
\be
m^2_H=9\frac{(g^{2}+g'^{2})^2+2g^4}{256\pi^2}\left(3M^2+\langle{\rho}\rangle^2\
ln\frac{(\langle{\rho}\rangle)^2}{M^2}-3\langle{\rho}\rangle^2\right)
\ee 
The mass of the $\rho$  field is computed to be $6.9$ Gev : perhaps too light to
be phenomenologically relevant as a standard Higgs field. It would be hard to identify
$\rho$ with a physical Higgs boson, the kind of which one expects to see at
the Large Hadron Collider. This mechanism of mass generation must therefore be
thought of as a toy model at this stage. However, it appears that LEP data
has not completely ruled out a very light Higgs boson with a mass less than 10
Gev \cite{Bae:2010hr} which may mediate elastic scattering between light
dark matter candidates.  

One may compare the ratio of vector boson masses to the Higgs mass generated
here with the corresponding result in the Coleman Weinberg framework \cite{Coleman:1973jx}.  The
ratios are not very different, even though we did not have to resort to
applying a `dimensional transmutation' here in order to cast the scalar mass
into a function entirely of the gauge couplings. This has been achieved here
quite naturally, since there is no scalar self-coupling in the theory at the
classical level.    

\section{Conclusion}

While the imaginative interpretation offered in the incipient work of
\cite{Chernodub:2008rz} appears to be untenable under radiative corrections, the
one loop effective Higgs potential generated in the theory has the prospect
of supplying the observed spectrum of weak vector boson masses, and possibly
also a Higgs mass; however, the latter is too low so as to preclude the theory,
if, as is commonly believed on the basis of LEP data, a Higgs boson is to be
detected at the Large Hadron 
Collider with a mass just above the 140 Gev range. The theory in this paper
has no self-coupling parameter, and the Higgs mass is completely determined by the
gauge couplings with appropriate choice of the renormalization scale. The
slightly disconcerting feature of this work is that the logarithm
$\log(\langle{\rho}\rangle)$ which features as an important dynamical factor
in the loop expansion, has a value of $O(1.67)$. So long as one restricts
oneself to low
orders of perturbation theory, this does not pose a serious problem at the
numerical level, even though theoretically it remains somewhat unsatisfactory.

An important issue is that of `naturalness' of the scalar sector of the
theory. Apart from the lacuna discussed above, this does {\it not} appear to
be an issue, since there is no scalar self coupling. The renormalization scale $M$ cannot
arbitrarily slide
to the GUT or the Planck scale without ruining the vector boson mass spectrum which is
extremely well determined experimentally. The scalar mass is then {\it constrained}
to be comparatively lower than the vector boson masses, and thus never
requires fine tuning of dimensionless parameters.  

The key question not addressed in this paper is of course the issue of fermion
masses. One could add to the Lagrange density (\ref{lagew}) fermionic gauge
and Yukawa coupling
terms where the Higgs vacuum expectation value produces fermion masses as in
the standard electroweak theory. The radiatively generated Higgs vev then can
be a source of fermion masses in the standard manner. However, in this case,
the Yukawa couplings control fermion masses and mixings much like in the standard
formulation, with no real economy in the size of the parameter space. The real challenge is to
produce the fermion masses dynamically at the quantum level. Chiral symmetry
prevents this from happening radiatiatively, so that the source of fermion
masses might have to be nonperturbative if a scenario akin to the model under
discussion ever becomes phenomenologically relevant.

\noindent {\bf Acknowledgment}: We would like to thank G. Bhattacharyya for
his initial comments about this problem.  We also thank Probir Roy and P. Byakti
for useful discussions on Standard Model phenomenology. S. B. would like to
thank R. Nandi for help with numerical techniques.

\end{document}